\documentclass[a4paper,11pt]{article}
\pdfoutput=1 

\usepackage{jheppub} 
\usepackage{neuralnetwork}
\usepackage{caption}
\usepackage{subcaption}
\usepackage{multicol}
\usepackage{graphicx} 
\usepackage{cancel}
\usepackage{grffile}
\usepackage{braket}
\usepackage{booktabs}
\usepackage{rotating}
\usepackage{multirow}
\usepackage{algorithm}
\usepackage{algpseudocode}
\usepackage{physics} 

\usepackage[normalem]{ulem}

\usepackage[T1]{fontenc} 

\preprint{JLAB-THY-23-3746, \, ADP-23-2/T1211}

\def\be{\begin{equation}}
\def\ee{\end{equation}}
\def\bea{\begin{eqnarray}}
\def\eea{\end{eqnarray}}

\author[a]{N.~T.~Hunt-Smith,}
\author[a,b]{W.~Melnitchouk,}
\author[b]{N.~Sato,}
\author[a]{A.~W.~Thomas,}
\author[a]{X.~G.~Wang,}
\author[a]{M.~J.~White}

\affiliation[a]{CSSM and ARC Centre of Excellence for Dark Matter Particle Physics, Department of Physics, University of Adelaide, Adelaide 5005, Australia}

\affiliation[b]{Jefferson Lab, Newport News, Virginia 23606, USA \\
        \vspace*{0.2cm}
        {\bf Jefferson Lab Angular Momentum (JAM) Collaboration
        \vspace*{0.2cm} }}

\emailAdd{nicholas.hunt-smith@adelaide.edu.au}
\emailAdd{wmelnitc@jlab.org}
\emailAdd{nsato@jlab.org}
\emailAdd{anthony.thomas@adelaide.edu.au}
\emailAdd{xuan-gong.wang@adelaide.edu.au}
\emailAdd{martin.white@adelaide.edu.au}

\title{\boldmath Global QCD analysis and dark photons}

\abstract{
We perform a global QCD analysis of high energy scattering data within the JAM Monte Carlo framework, including a coupling to a dark photon that augments the Standard Model (SM) electroweak coupling via kinetic mixing with the hypercharge $B$ boson. 
We first set limits on the dark photon mass and mixing parameter assuming that the SM is the true theory of Nature, taking into account also the effect on $g-2$ of the muon. 
If instead we entertain the possibility that the dark photon may play a role in deep-inelastic scattering (DIS), we find that the best fit is preferred over the SM at 6.5$\sigma$, even after accounting for missing higher order uncertainties. 
The improvement in $\chi^2$ with the dark photon is stable against all the tests we have applied, with the improvements in the theoretical predictions spread across a wide range of $x$ and~$Q^2$. 
The largest improvement corresponds to the fixed target and HERA DIS data, while the best fit yields a value of $g-2$ which significantly reduces the disagreement with the latest experimental determination. 
}

\begin{document}
\maketitle
\flushbottom

\section{Introduction}
\label{sec:introduction}
Despite the enormous success of the Standard Model (SM) of nuclear and particle physics, it remains an incomplete theory, not least because of its inability to explain dark matter.
One relatively simple addition that could be made to at least provide a portal to the dark sector would be to introduce a new  massive U(1) gauge boson~\cite{Fayet:1980ad, Fayet:1980rr, Holdom:1985ag}, referred to as the dark photon.
Here the dark photon is chosen to mix kinetically with the SM hypercharge boson, requiring the additional Lagrangian terms~\cite{Okun:1982xi},
\begin{equation}
\mathcal{L}  \supset - \frac{1}{4} F'_{\mu\nu} F'^{\mu\nu} + \frac{m^2_{A'}}{2} A'_{\mu} A'^{\mu} + \frac{\epsilon}{2 \cos\theta_W} F'_{\mu\nu} B^{\mu\nu} \, ,
\end{equation}
where $A'$ denotes the unmixed version of the dark photon. 
The parameter $\epsilon$ describes the degree of mixing between the dark photon and the $B$ boson of the standard electroweak theory, $\theta_W$ is the Weinberg angle, and $F'_{\mu \nu}$ is the dark photon field strength tensor.
After electroweak symmetry breaking and diagonalizing the kinetic terms and gauge boson masses, three physical vectors remain which couple to the SM fermions: the massless photon~$\gamma$, the massive $Z$ boson, and the physical dark photon, labelled~$A_D$.

Many accelerator-based searches for the dark photon have been undertaken~\cite{BaBar:2017tiz, Banerjee:2019pds, LHCb:2019vmc, CMS:2019buh}, with none observing a signal to date.
Large regions of parameter space with mixing parameter $\epsilon > 10^{-3}$ in both light and heavy mass regions have been ruled out~\cite{Graham:2021ggy}, with a few gaps, which are significant in the light of our results, associated with the production of resonances, such as the $J/\psi$ and its excited states. 
Further competitive constraints have recently been placed on the dark photon from ``decay-agnostic" (independent of decay modes or production mechanism) processes, such as the muon $g-2$ anomaly~\cite{Pospelov:2008zw, Davoudiasl:2012qa}, the electroweak precision observables (EWPO)~\cite{Hook:2010tw, Curtin:2014cca,Loizos:2023xbj}, $e^\pm p$ deep-inelastic scattering (DIS)~\cite{Kribs:2020vyk, Thomas:2021lub, Yan:2022npz}, parity-violating electron scattering~\cite{Thomas:2022qhj, Thomas:2022gib}, rare kaon and $B$-meson decays~\cite{Davoudiasl:2012ag, Wang:2023css}, and high-luminosity LHC projections~\cite{McCullough:2022hzr}.

\begin{figure}[t]
\centering
\includegraphics[scale = 0.45]{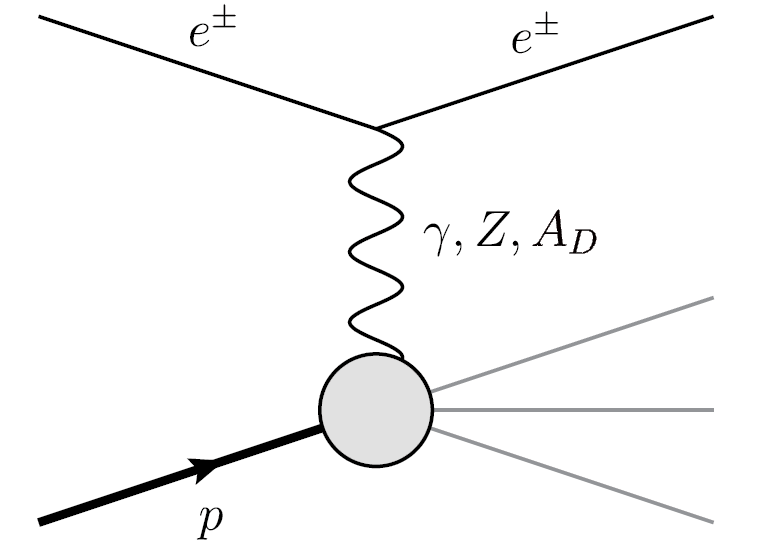}
\caption{Kinetic mixing of the dark photon $A_D$ with SM mediators $\gamma$ and $Z$ in $e^\pm p$ DIS~\cite{Kribs:2020vyk}.}
\label{fig:feynmann}
\end{figure}

The dark photon contributes to DIS processes coherently along with photon and $Z$ boson exchange, as illustrated in Fig.~\ref{fig:feynmann}. 
It has also been shown to be necessary to simultaneously extract the parton distribution functions (PDFs) from the data when incorporating beyond the SM (BSM) physics into proton structure~\cite{Carrazza:2019sec, Kassabov:2023hbm, Gao:2022srd, Iranipour:2022iak, Greljo:2021kvv}.
A recent exploratory study~\cite{Thomas:2021lub} included an extraction of the PDFs alongside the dark photon contribution, although that analysis limited itself to a subset of the existing HERA and BCDMS data and employed only a basic leading order (LO) analysis of DIS hard scattering to extract PDFs.

Here we report the first global QCD analysis including a dark photon within the JAM next-to-leading order (NLO) analysis framework. 
This approach employs modern Monte Carlo techniques and state-of-the-art uncertainty quantification, taking into account power corrections and nuclear effects in the case of the deuteron~\cite{Cocuzza:2021cbi}. 
We perform two different analyses in this paper.
In the first, we assume that the SM is the correct theory of Nature and present an exclusion limit in the plane of the dark photon mass and mixing parameters.
In the second, we allow for the possibility that the dark photon may improve the likelihood for experimental observables relative to the SM, and simultaneously determine the optimum set of PDFs, as well as the preferred dark photon parameters. In scanning the parameter space, we consider all possible values for the dark photon mass, and do not make any assumptions about its numerical value. 
We then perform a hypothesis test against the SM to quantify the improvement at the best fit point.
Having identified a preference for the dark photon model, we test the stability of this conclusion against the inclusion of power corrections, an increase in the lower cutoff in $Q^2$ from $m_c^2$ to 10~GeV$^2$, and an estimate of missing higher order uncertainties following Ref.~\cite{NNPDF:2019ubu}.

In Sec.~\ref{sec:background} we begin by reviewing the modifications to the $F_2$ proton structure function with the addition of a dark photon contribution.
Section~\ref{sec:methodology} outlines the methodological details of the global QCD analysis performed within the JAM framework.
Our results are presented in Sec.~\ref{sec:results}, first placing exclusion limits on the dark photon parameters, and then exploring possible improvements of the global fit with the inclusion of the dark photon. 
Finally, in Sec.~\ref{sec:outlook} we summarize our conclusions and discuss the implications and future extensions of this work.

\section{Dark photon background}
\label{sec:background}

The $F_2$ and $F_3$ proton structure functions, including the electroweak and dark photon contributions, are given by~\cite{Kribs:2020vyk}
\begin{equation}
    \begin{aligned}
        & \widetilde{F}{}_{2} = \sum_{i,j = \gamma,Z,A_D} \kappa_i \kappa_j F^{ij}_2, \\
        & \widetilde{F}{}_{3} = \sum_{i,j = \gamma,Z,A_D} \kappa_i \kappa_j F^{ij}_3,
    \end{aligned}
\end{equation}
where $\kappa_i = Q^2/(Q^2 + m_i^2)$.
At LO in the strong coupling $\alpha_s$, one has
\begin{equation}
    \begin{aligned}
        & F^{ij}_2 = \sum_q (C^v_{i,e} C^v_{j,e} + C^a_{i,e} C^a_{j,e})(C^v_{i,q} C^v_{j,q} + C^a_{i,q} C^a_{j,q})\, x f_q, \\
        & F^{ij}_3 = \sum_q (C^v_{i,e} C^a_{j,e} + C^a_{i,e} C^v_{j,e})(C^v_{i,q} C^a_{j,q} + C^a_{i,q} C^v_{j,q})\, x f_q,
    \end{aligned}
\end{equation}
where $x$ is the parton momentum fraction, and $f_q$ is the PDF for quark flavor $q$ in the proton.
The vector and axial vector couplings to the electron and quarks for the photon are
\begin{equation}
    \{C^v_{\gamma,e},C^v_{\gamma,u},C^v_{\gamma,d}\} 
    = \left\{-1,\, \frac{2}{3},\, -\frac{1}{3}\right\},\hspace{5mm} C^a_\gamma = 0,
\end{equation}
while for the unmixed $Z$ boson the couplings are
\begin{equation}
    \overline{C}^v_Z \sin{2\theta_W} = T^f_3 - 2 q_f \sin^2{\theta_W}, \hspace{3mm}
    \overline{C}^a_Z \sin{2\theta_W} = T^f_3,
\end{equation}
where $T^f_3$ is the third component of the weak isospin, and $q_f$ the electric charge.
%
%
After diagonalizing the mixing term through field redefinitions, the couplings of the physical $Z$ and $A_D$ to SM particles are given by~\cite{Kribs:2020vyk}
\begin{equation}
    \begin{aligned}
        & C^v_Z = (\cos{\alpha} - \epsilon_W \sin{\alpha})\, \overline{C}^v_Z
        + \epsilon_W \sin{\alpha}\cot{\theta_W}\, C^v_\gamma, \\
        & C^a_Z = (\cos{\alpha} - \epsilon_W \sin{\alpha})\, \overline{C}^a_Z,
    \end{aligned}
\end{equation}
and
\begin{equation}
    \begin{aligned}
        & C^v_{A_D} = -(\sin{\alpha} + \epsilon_W \cos{\alpha})\, \overline{C}^v_Z + \epsilon_W \cos{\alpha}\cot{\theta_W} C^v_\gamma, \\
        & C^a_{A_D} = -(\sin{\alpha} + \epsilon_W \cos{\alpha})\, \overline{C}^a_Z.
    \end{aligned}
\end{equation}
Here $\alpha$ is the ${\bar Z}$--$A'$ mixing angle,
\begin{equation}
    \begin{aligned}
        \tan{\alpha} =\, &\frac{1}{2 \epsilon_W} \Big[1-\epsilon^2_W - \rho^2 \\
        & -\, \text{sign}\, (1-\rho^2)\sqrt{4 \epsilon^2_W + (1-\epsilon^2_W - \rho^2)^2}\,\Big] \, ,
    \end{aligned}
\end{equation}
with $\epsilon_W$ given in terms of the free mixing parameter~$\epsilon$,
\begin{equation}
    \epsilon_W = \frac{\epsilon \tan{\theta_W}}{\sqrt{1-\epsilon^2/\cos^2{\theta_W}}} \, ,
\end{equation}
and $\rho$ is defined by
\begin{equation}
    \rho = \frac{  m_{A'}/m_{\bar{Z}} }
    {\sqrt{1-\epsilon^2/\cos^2{\theta_W}}} \, .
\end{equation}
The physical masses of the $Z$ boson and dark photon then become
\begin{equation}
    \begin{aligned}
        m^2_{Z,A_D} = &\frac{m^2_{\bar{Z}}}{2} \Big[1+\epsilon^2_W + \rho^2 \\
        & \pm\, \text{sign}\, (1-\rho^2)\sqrt{(1+\epsilon^2_W + \rho^2)^2 -4\rho^2}\,\Big] \, ,
    \end{aligned}
\end{equation}
with $\bar{Z}$ the unmixed version of the SM neutral weak boson.
In our analysis we include the two dark parameters, $\epsilon$ and the mass $m_{A_D}$, amongst the fitting parameters.

\section{Methodology}
\label{sec:methodology}
Our baseline study uses the JAM QCD analysis framework, employing Monte Carlo sampling and fixed order perturbation theory at NLO accuracy in the QCD coupling, $\alpha_s$.
To accurately characterize the PDFs and their uncertainties, as well as the dark photon parameters, the global analysis was performed 200 times using data resampling, repeatedly fitting to data that were distorted by Gaussian shifts within their quoted uncertainties via $\chi^2$ minimization.
The resulting replica parameter sets approximate Bayesian samples of the posterior, from which confidence levels for the PDFs and dark parameters may be estimated~\cite{Hunt-Smith:2022ugn}.
Starting from the replicas of the previous JAM fit~\cite{Cocuzza:2021cbi}, our global analysis incorporates the JAM multi-step strategy, whereby data are sequentially added one step at a time, as an efficient way of locating the minimum $\chi^2$ for a given shuffling of the data. 
This in turn ensures we obtain a good approximation to the posterior over the 200 replicas.

For inclusive DIS, QCD factorization theorems allow us to write the $F_2$ and $F_L$ structure functions as sums of convolutions of hard scattering functions, nonperturbative quark and gluon PDFs, and higher twist power corrections,
\begin{subequations}
\label{eq:F2LDIS}
\begin{eqnarray}
    F_2(x,Q^2) &=& \Big( \sum_q e^2_q [C_{2q} \otimes q^+] + [C_{2g} \otimes g] \Big)(x,q^2)\, 
    \Big( 1+ \frac{C^{\rm HT}_2(x)}{Q^2}\Big),
\label{eq:F2DIS}
    \\
    F_L(x,Q^2) &=& \Big( \sum_q e^2_q [C_{Lq} \otimes q^+] + [C_{Lg} \otimes g \Big)(x,q^2)\,
    \Big( 1+ \frac{C^{\rm HT}_L(x)}{Q^2}\Big),
\label{eq:FLDIS}
\end{eqnarray}
\end{subequations}
where $q_N^+ = q_N + \overline{q}_N$, and the hard scattering coefficient functions $C_{ij}$ ($i=2,L$; $j=q,g$) are computed in fixed order perturbation theory at NLO accuracy in the QCD coupling, following the ``track B'' approach of Refs.~\cite{Collins:2021vke, Collins:2011zzd}.
At this order, the leading contributions to $C_{2j}$ are of order ${\cal O}(\alpha_s^0)$, while $C_{Lj}$ enters at order ${\cal O}(\alpha_s)$.
We parametrize the PDFs at the input scale $Q^2_0$ using the standard form,
\begin{equation}
    f(x,Q^2_0) = N x^\alpha (1 - x)^\beta (1 + \gamma \sqrt x + \eta x),
\end{equation}
so that each individual fit determines 30 free proton PDF parameters. 
For the heavy quarks, in our analysis we use the zero mass variable flavor scheme.
The $C_i^{\rm HT}(x)$ higher twist coefficients can in principle be different for protons and neutrons, and are determined phenomenologically from low-$Q^2$ data by introducing a further 6 free parameters to be fitted. 
In this analysis, we assume a multiplicative {\it ansatz} for the coefficients $C_i^{\rm HT}$, with the higher twist parameters taken to be the same for the longitudinal structure function $F_L$ as for $F_2$. 
The definitions of the structure functions in Eqs.~(\ref{eq:F2LDIS}) also contain target mass corrections implemented within the collinear factorization as described in Refs.~\cite{Aivazis:1993kh, Moffat:2019qll}.

In the nuclear impulse approximation at $x \gg 0$, the scattering takes place incoherently from individual (off-shell) nucleon contributions in the nucleus.
As such, we can write the nuclear PDFs as a sum of on-shell and off-shell contributions ~\cite{Melnitchouk:1993nk, Melnitchouk:1994rv, Kulagin:1994fz},
\begin{equation}
    q_A(x,Q^2) = \sum_N q_{N/A}(x,Q^2) = \sum_N \big[q^{\rm (on)}_{N/A} + q^{\rm (off)}_{N/A}\big](x,Q^2),
\end{equation}
where $q_{N/A}$ refers to the PDF of a quark $q$ in a nucleon $N$, as modified within a nucleus $A$.
We introduce 6 more free parameters to account for these off-shell contributions, which for our chosen datasets only takes the form of small corrections to the deuterium data. Overall, this results in a total of 42 free PDF parameters in our baseline analysis.

The DIS datasets included in this analysis consist of fixed target proton and deuteron DIS data from SLAC~\cite{Whitlow:1991uw}, BCDMS~\cite{BCDMS:1989qop}, and NMC~\cite{NewMuon:1996fwh, NewMuon:1996uwk}, and collider data from HERA~\cite{H1:2015ubc}. 
For the fixed target DIS data the $F_2$ structure function is provided, while for the HERA data the reduced cross sections are presented.
At lower $x$ and $Q^2$ values, the longitudinal contribution to the cross section should be taken into account, while in the kinematics of the HERA data included in our analysis the $F_L$ contribution to the reduced cross sections, computed using Eq.~(\ref{eq:FLDIS}), is negligible.
However, we still include the $F_L$ contribution, especially in a precise determination of the dark photon parameters whose effect is also expected to be small. 
For all DIS datasets, a $Q^2$ cut of 1.69~GeV$^2$ and a $W^2$ cut of 10~GeV$^2$ were employed, as in Ref.~\cite{Cocuzza:2021cbi}.
Other datasets that were added to constrain the PDF parameters include $pp$ and $pd$ Drell-Yan data from the Fermilab NuSea~\cite{NuSea:2001idv} and SeaQuest~\cite{SeaQuest:2021zxb} experiments, $Z$-boson rapidity data~\cite{CDF:2010vek, D0:2007djv}, $W$-boson asymmetry data~\cite{CDF:2009cjw, D0:2013lql}, and jet production data from $p\bar p$ collisions at the Tevatron~\cite{CDF:2007bvv, D0:2011jpq}.

The modifications required to include the effects of the dark photon were then added to the underlying JAM theory, allowing the two additional dark parameters, $m_{A_D}$ and $\epsilon$, to be fitted alongside the PDF parameters. 
After repeating the global fits, we then compared the 200 replicas obtained with and without the dark photon. 
In global QCD analyses the agreement between the fitted results and the data is assessed through the reduced $\chi^2$, which computes the difference between the theory predictions averaged over the replicas and the actual data.

Because of its particular sensitivity to a dark photon, we  included an additional contribution to the total $\chi^2$ corresponding to the value of $g - 2$ for the muon, for which there is currently an anomaly~\cite{Muong-2:2021ojo, Muong-2:2006rrc}.
The SM predicts a result for $g - 2$ that differs from the experimental measurement by $(251 \pm 59) \times 10^{-11}$ (which has a significance of 4.2$\sigma$). 
While there are indications from lattice QCD~\cite{Borsanyi:2020mff} for a somewhat higher SM value, there is no consensus on this in the lattice community, nor a successful reconciliation with the dispersion relation method.
We therefore use the value reported in Ref.~\cite{Aoyama:2020ynm} and quoted in Refs.~\cite{Muong-2:2021ojo, Muong-2:2006rrc}. 
In the dark photon scenario the correction to $g - 2$ scales as $m_{\ell}^2 \, \epsilon^2/m^2_{A_D}$ (with $m_{\ell}$ the lepton mass)~\cite{Pospelov:2008zw}, meaning that the experimental measurement can act as a powerful constraint on the dark photon parameters.

\section{Global QCD analysis results}
\label{sec:results}

\begin{figure}[t]
\centering
\includegraphics[scale = 0.6]{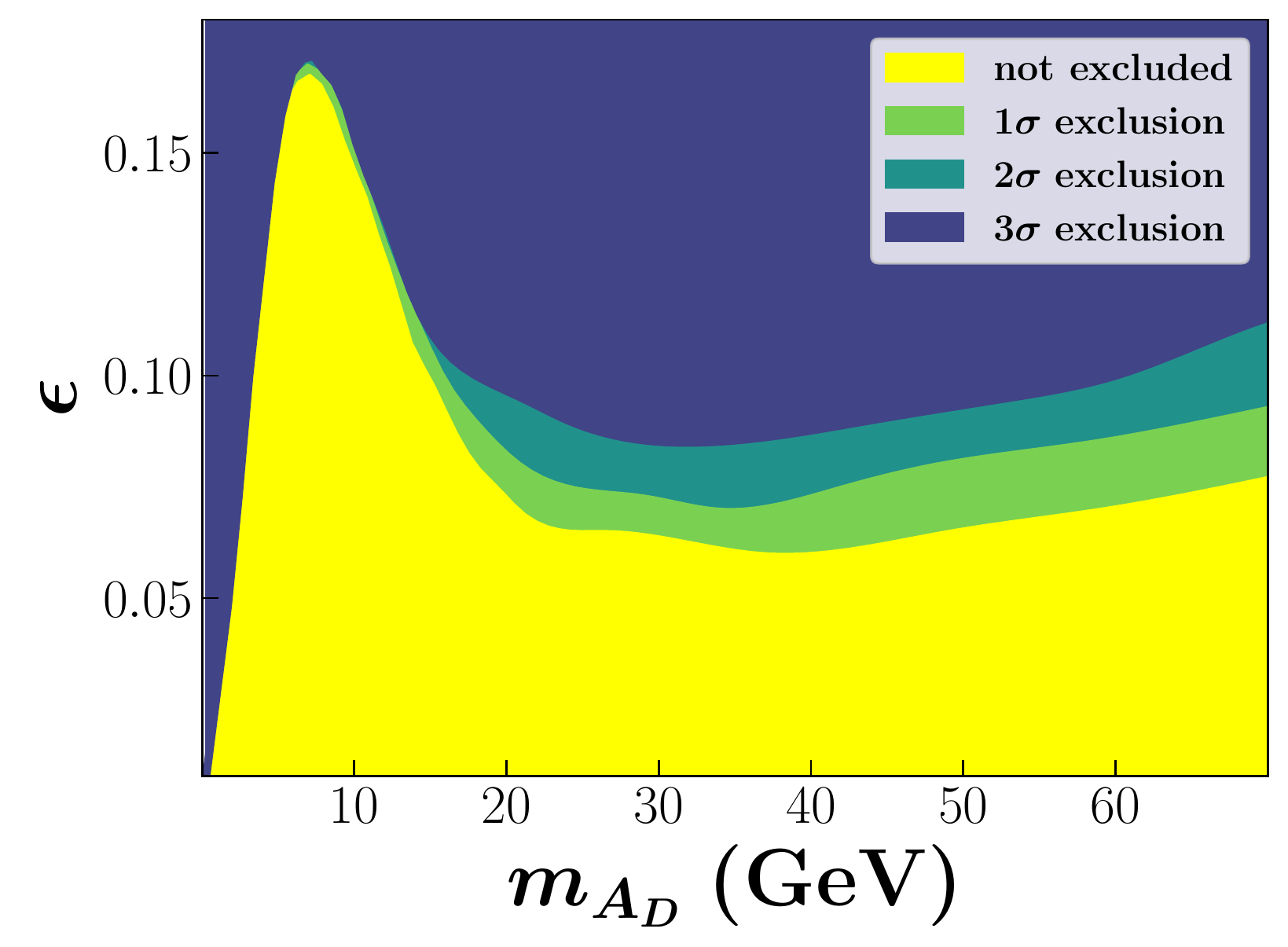}
\caption{The excluded region in the dark photon parameter space, assuming that the SM is the correct theory of Nature.}
\label{fig:exclusion}
\end{figure}

We first explore the region of parameter space where a $\chi^2$ analysis forbids the existence of the dark photon, assuming that the SM is the true theory. 
This is accomplished by not allowing the $\chi^2$ to be lower than that obtained with the best fit SM hypothesis, and is similar to the ``capped likelihood'' treatment followed in Ref.~\cite{GAMBIT:2018gjo}. 
The excluded region is shown in Fig.~\ref{fig:exclusion}, where we see that, for $m_{A_D}>20$~GeV, $\epsilon$ must be less than $\approx 0.08$.
In the region around 10~GeV, the mixing parameter $\epsilon$ is allowed to be notably larger due to the dark photon fitting anomalies in some datasets, whilst being penalized by others sufficiently to give the same quality of fit as the SM. 
This weakens the excluded range of $\epsilon$.

We also note that the exclusion limits over the whole mass range for $\epsilon$ in Fig.~\ref{fig:exclusion} are somewhat weaker than those reported in earlier analyses of DIS data \cite{Kribs:2020vyk, Thomas:2021lub}.
We first observe that the strong constraints derived in Ref.~\cite{Kribs:2020vyk} are mainly driven by:
(i) the input PDFs were fixed at the values resulting from the best fit results of a HERA analysis, leaving no room for the potential effects of new physics on the extracted PDFs; 
(ii) the uncertainties in the reduced neutral current (NC) cross section $\sigma^{\rm NC}_{\rm red}$ were assumed to be at most 0.3\%--0.4\%, which is approximately 4 times smaller than those of the raw data~\cite{H1:2015ubc}. 
Moreover, the longitudinal structure functions were neglected in Ref.~\cite{Kribs:2020vyk}, which is a good approximation only when $y=Q^2/xs$ is small. 
In our work, we have analyzed the original HERA data including both the transverse and longitudinal structure functions, and have allowed the PDFs to adjust to the presence of the dark photon.
Compared with Ref.~\cite{Thomas:2021lub}, the dataset used here is considerably larger and the restrictions imposed on the PDF parameters in that work have been removed.

\begin{table}[b]
\caption{Comparison of the $\chi^2$ values per degree of freedom, $\chi^2_{\rm dof}$, with (``dark'') and without (``base'') dark photon modifications for various datasets, taking into account a correction for missing higher order uncertainties (see text).}
\begin{center}
\begin{tabular}{p{8em}|p{8em}|p{8em}|p{8em}}
    ~reaction & ~$\chi^2_{\rm dof}({\rm dark})$~ & ~$\chi^2_{\rm dof}({\rm base)}$~ & ~$N_{\rm dof}$~ \\
    \hline 
    ~fixed target DIS & 1.01 & 1.05 & 1495~~\\
    ~HERA NC          & 1.02 & 1.03 & 1104~~\\
    ~HERA CC          & 1.13 & 1.18 & 81~ \\
    ~Drell-Yan        & 1.18 & 1.16 & 205~ \\ 
    ~$Z$ rapidity     & 1.08 & 1.05 & 56~ \\ 
    ~$W$ asymmetry    & 1.04 & 1.07 & 97~ \\ 
    ~jets             & 1.16 & 1.15 & 200~ \\ 
    \hline
    ~\bf{total} & 1.03 & 1.05 & 3283~ \\
\end{tabular}
\label{table:chi2}
\end{center}
\end{table}

%
\begin{figure}[t]
\centering
\includegraphics[scale = 0.6]{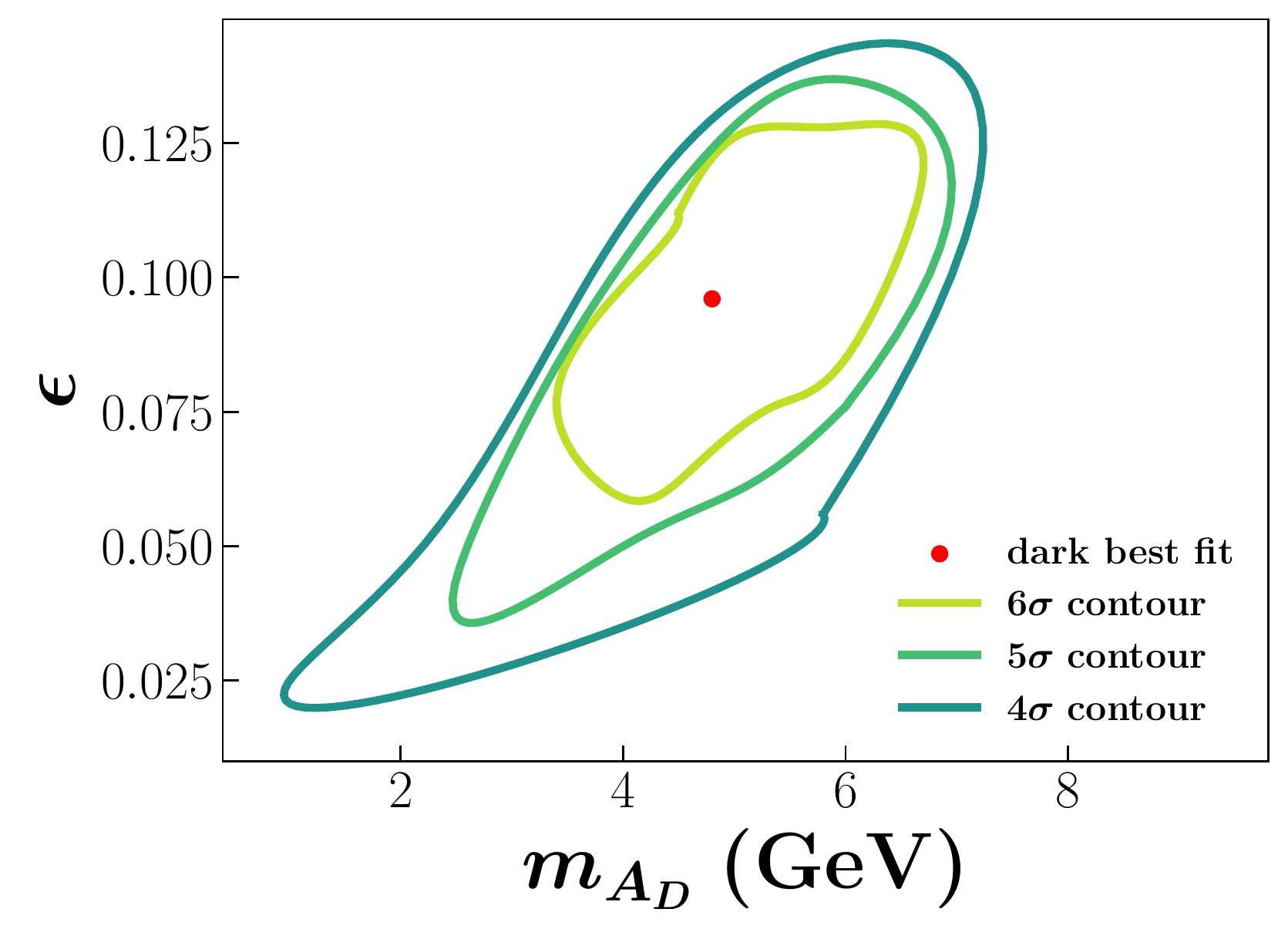}
\caption{Results of an hypothesis test for the likelihood that the SM is the correct theory to describe this data, compared with the case where a dark photon is included. The hypothesis that the SM is the correct theory is excluded at 6.5$\sigma$ for the best dark photon fit at the red point.}
\label{fig:hypothesis}
\end{figure}

Next, we present results of a fit where the dark photon is allowed to improve the fit to the experimental observables.
A comparison of the $\chi^2$ per degree of freedom between the baseline average theory and the dark photon average theory for the different datasets included is given in Table~\ref{table:chi2}.
There is a clear improvement in the reduced $\chi^2$ for the fixed target DIS and HERA datasets with the inclusion of the dark photon. This improvement was essentially independent of the number of replicas considered.
All the datasets, other than the Drell-Yan, $Z$ rapidity and jet data, exhibit a slightly lower $\chi^2$ with the dark photon, although no modifications were made to the underlying theory for those observables. In fact, electroweak corrections to the Drell-Yan cross sections are small and generally omitted in global PDF analyses. However, as a check, we did include the dark photon contribution, which yielded a negligible improvement in $\chi^2$ of 0.003 per degree of freedom for that dataset.
The sheer number of DIS data points means that the total $\chi^2$ shows a substantial improvement.
The best fit including the dark photon also reduces the difference in the anomalous value of $g \, - \, 2$ between the SM prediction and experiment from $251 \times 10^{-11}$ to $91 \times 10^{-11}$, which in turn reduces the statistical significance of the discrepancy in that observable from 4.2$\sigma$ to 1.5$\sigma$.

In order to assess the statistical significance of this $\chi^2$ improvement, we performed an hypothesis test of the dark photon hypothesis against the SM-only hypothesis. 
Since these hypotheses are nested we can make use of the log-likelihood test statistic, defined as
\begin{equation}
    t = -2\, \big[ \ln\mathcal{L}(\hat{\nu},m_{A_D},\epsilon) - \ln\mathcal{L}(\hat{\nu},\hat{m}_{A_D},\hat{\epsilon}) \big],
\end{equation}
where $\mathcal{L}(\hat{\nu},m_{A_D},\epsilon)$ is the likelihood for the SM-only hypothesis (or ``baseline''), meaning a choice of $m_{A_D} = \infty$ and $\epsilon = 0$, maximized over the nuisance PDF parameters~$\hat{\nu}$. 
The likelihood $\mathcal{L}(\hat{\nu},\hat{m}_{A_D},\hat{\epsilon})$ is then maximized over the dark parameters, in addition to the PDF parameters.
Using our choice of likelihood function $\mathcal{L} = \exp\left(-\frac12 \chi^2\right)$, the test statistic reduces to the difference of the $\chi^2$ values between the baseline and dark best fits, $t = \chi^2_{\rm base} - \chi^2_{\rm dark}$. 
The results of the fit are shown in Fig.~\ref{fig:hypothesis}, and applying Wilks' theorem to the $t$ test statistic, the difference in total $\chi^2$ for the best fit corresponds to a $p$-value of $\sim 3.8 \times 10^{-11}$ or $\sim 6.5\sigma$. 
We checked the validity of applying Wilks' theorem to our results, confirming that we do indeed have a $\chi^2$-distributed likelihood ratio test statistic. 
This implies that the quoted $p$-value should be a good approximation.
\begin{figure}[b]
\centering
\includegraphics[scale = 0.6]{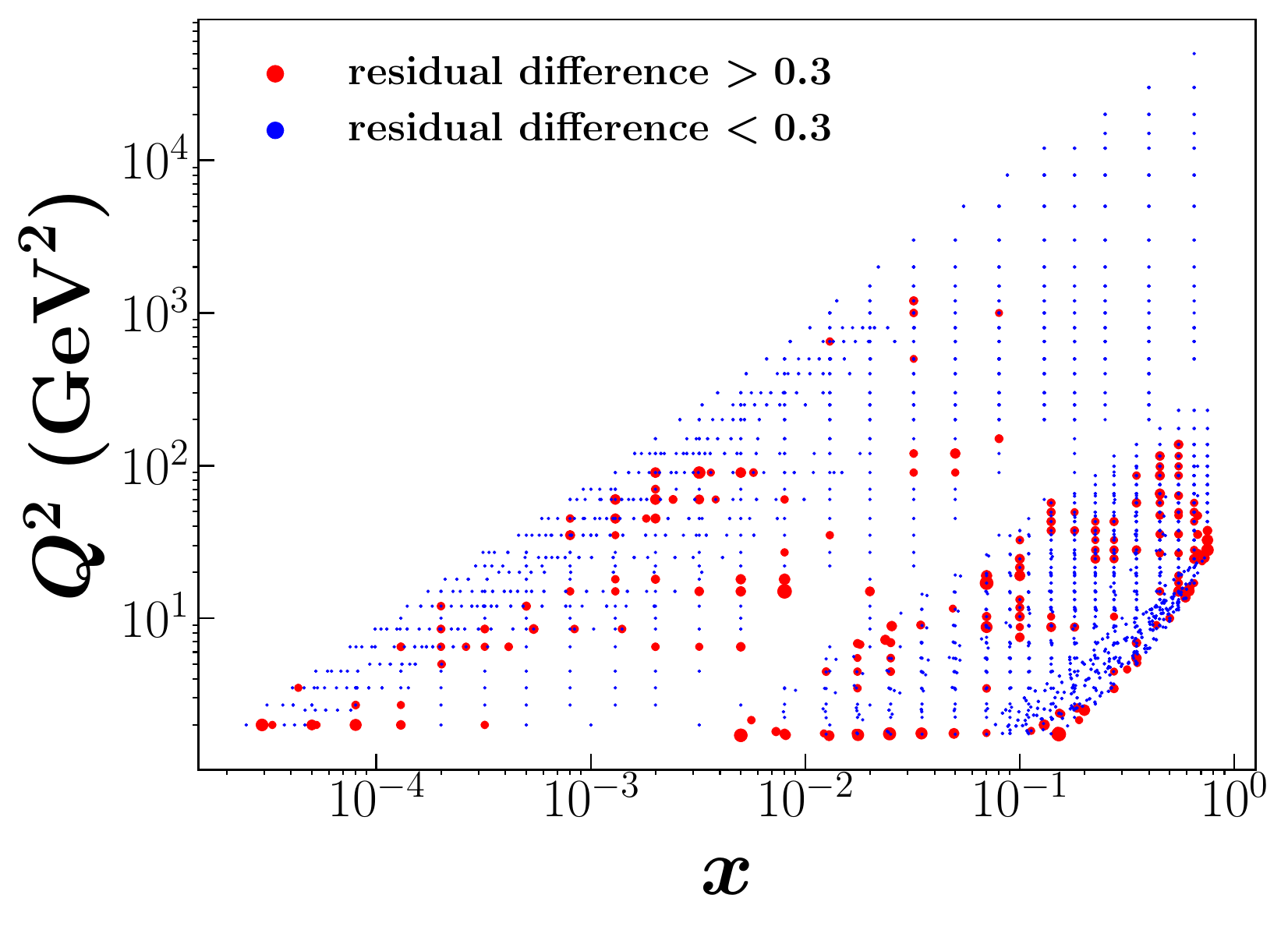}
\caption{Points in DIS kinematics $(Q^2,x)$ exhibiting the greatest improvement in residuals ($> 0.3$ in red, $< 0.3$ in blue) between dark and baseline average theory.}
\label{fig:residuals}
\end{figure}

There is no localized region of kinematics which contains the greatest difference in residuals, as seen in Fig.~\ref{fig:residuals}.
The marker size of the data points corresponds to an improvement in the absolute value of the residual less than 0.3 set to a very small size and shown in blue, while an improvement larger than this is shown by a larger point in red.
The improvement associated with the inclusion of the dark photon covers most of the range of $Q^2$ values for each given $x$, as shown by the spread of red points in Fig.~\ref{fig:residuals}. 
This indicates that the improvement in $\chi^2$ seen with our dark photon modifications is unlikely to be  explained by some missing element in the QCD theory that contributes via a specific region of $Q^2$ or $x$.

The inclusion of next-to-next-to-leading order (NNLO) corrections has been shown to improve the fit for HERA data~\cite{Harland-Lang:2016yfn} to an extent that could be comparable with that reported here by including the dark photon modifications. 
As a quantitative test of the potential effects of higher order corrections, or so-called missing higher order uncertainties (MHOU), we follow the procedure developed by the NNPDF Collaboration~\cite{NNPDF:2019ubu}.
That is, we add in quadrature to the experimental uncertainty on each data point the largest variation in the theoretical prediction obtained by varying the $Q^2$ for that point by a factor of anywhere between 0.5 and 2.0. 
This reduced the total $\chi^2$ by roughly 250, but still left a very significant improvement when the dark photon was added.
The results for the $\chi^2$ per degree of freedom shown in Table~\ref{table:chi2} include this additional uncertainty.

As a test of the effect of power corrections on the improvement in $\chi^2$ associated with the dark photon, we repeated the analysis with power corrections turned off.
The total $\chi^2$ increased by around 16, but the reduction associated with the dark photon was unchanged. 
To check the effects of data at relatively low-$Q^2$, we also repeated the analysis with a lower bound of 10~GeV$^2$ instead of 1.69~GeV$^2$.
This reduced the number of data points significantly but the improvement in $\chi^2$ was still such as to favor the existence of the dark photon by 5$\sigma$.

\begin{figure}[t]
\centering
\includegraphics[scale = 0.3]{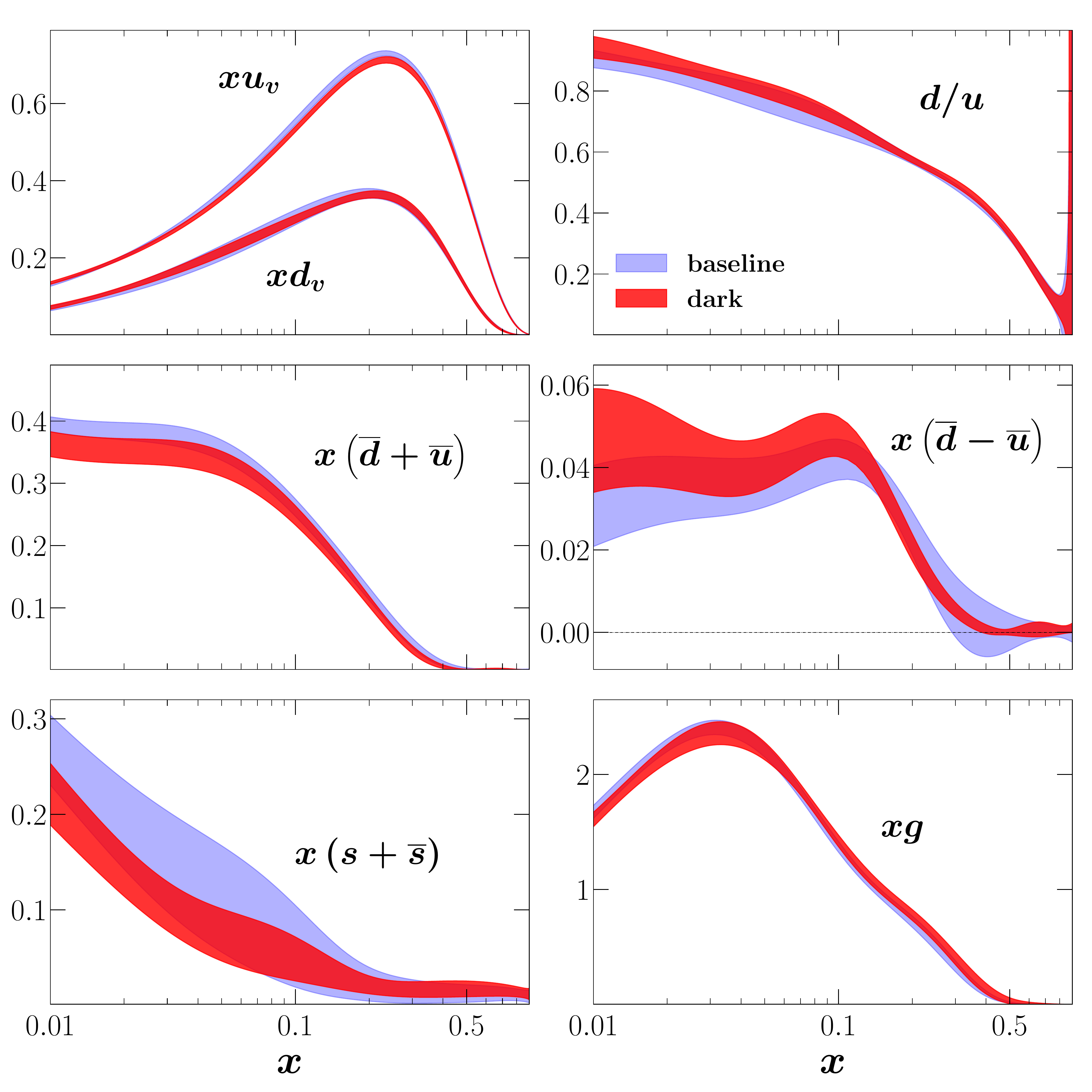}
\caption{A comparison of the PDF replicas with (``dark'', red bands) and without (``baseline'', blue bands) the dark photon modifications at the input scale, $Q^2=1.69$~GeV$^2$.}
\label{fig:pdfs}
\end{figure}

In Fig.~\ref{fig:pdfs} we compare the PDFs from the dark photon and baseline replicas.
The dark PDFs overlap with the baseline PDF uncertainty bands most of the time, particularly for the valence quark distributions that are best constrained by existing data. 
We also found that both the dark and baseline PDFs are similar to the most recent JAM22 unpolarized PDF results~\cite{Cocuzza:2021cbi}, up to some small differences in the $W^2$ cut used in that analysis. This suggests that the dark photon results are indeed consistent, within uncertainties, with existing PDF analyses, albeit with small but potentially interesting shifts in the central values.

\begin{figure}[t]
\centering
\includegraphics[scale = 0.6]{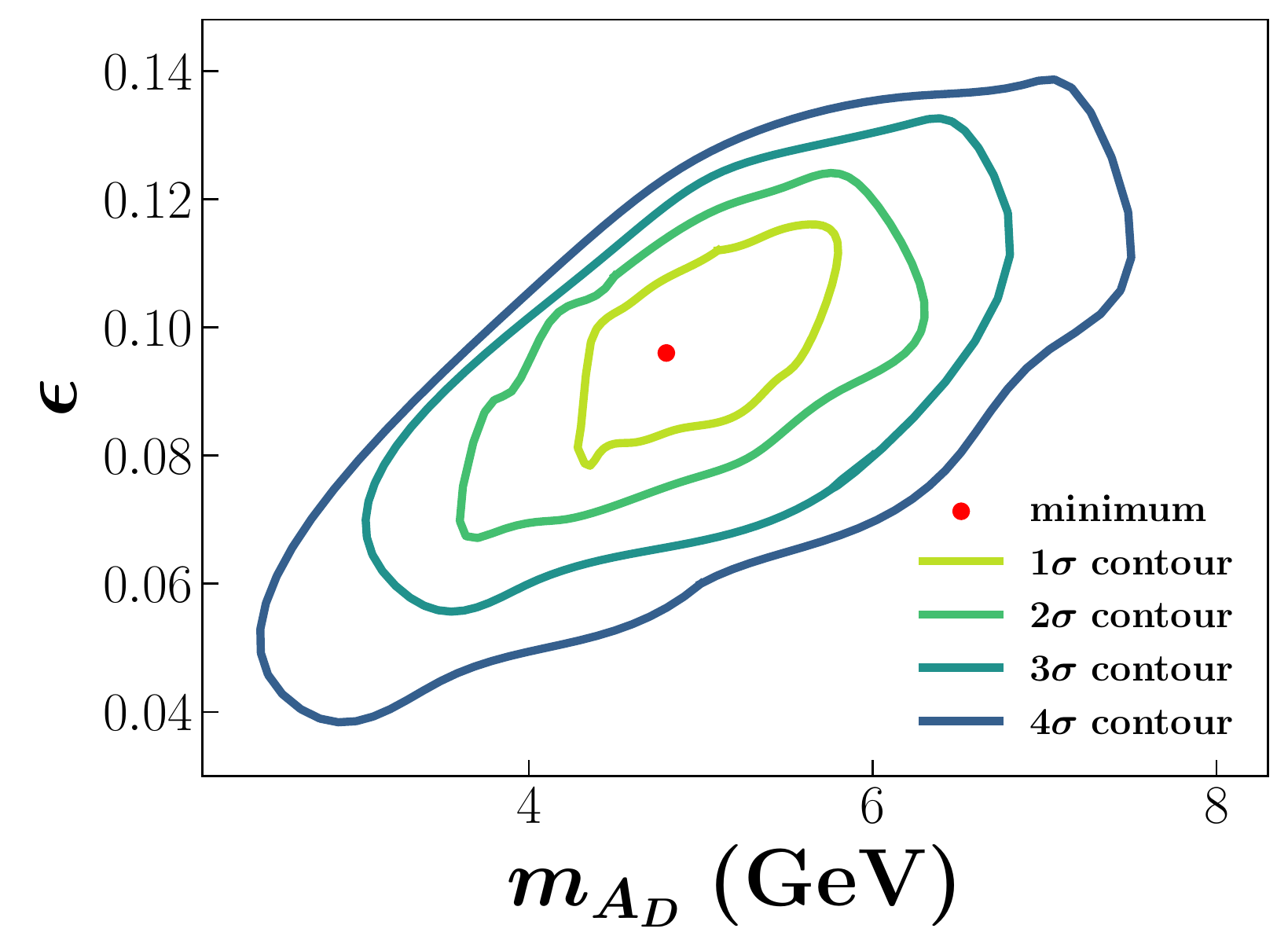}
\caption{Profile likelihood for the dark photon mass $m_{A_D}$ versus the mixing parameter $\epsilon$.}
\label{fig:contours} 
\end{figure}

Figure~\ref{fig:contours} illustrates the profile likelihood in the $\epsilon$--$m_{A_D}$ plane, with contours indicating 1$\sigma$, 2$\sigma$, 3$\sigma$ and 4$\sigma$ levels.
The ranges of the dark parameters favored by the data give a dark photon mass $m_{A_D} = 4.8^{+1.5}_{-1.2}$~GeV and mixing parameter $\epsilon = 0.096^{+0.028}_{-0.028}$ at the 95\% confidence level.
This value of $\epsilon$ appears somewhat high, as $\epsilon > 10^{-3}$ has been largely excluded for a wide range of mass values~\cite{Graham:2021ggy}.
On the other hand, there do exist regions of parameter space within this mass range which cannot be excluded because of the presence of other particles, notably the $J/\psi$ and its excited states between $\approx 3$~GeV and 4.5~GeV.

The preferred value for the dark photon mass has an uncertainty range that is wide enough to encompass these unconstrained regions, meaning that our fitted values for the dark parameters are not inconsistent with existing experiments. 
In fact, the improvement in $\chi^2$ is so substantial that if we perform the hypothesis test at the point $\epsilon = 0.05$ and mass $m_{A_D} = 3$~GeV, the SM-only hypothesis is still excluded with a significance above $5 \sigma$. 

Finally, it is interesting to comment on the parameters preferred in this analysis in comparison with other existing constraints. 
Constraints from EWPO~\cite{Hook:2010tw, Curtin:2014cca} typically require a mixing parameter for a dark photon somewhat below our preferred range, in the relevant mass region. 
This needs a careful reanalysis taking into account allowed variations in key SM parameters, as well as new information on the $W$-boson mass~\cite{CDF:2022hxs, ATLAS-CONF-2023-004}. 
In the case of $g - 2$ for the electron our results give no significant change to the current satisfactory situation of experiment versus theory~\cite{Morel:2020dww, Fan:2022eto}, with the uncertainty in $\alpha_{\rm em}$ being dominant.

\section{Outlook}
\label{sec:outlook}

In summary, we have mapped out the region of dark photon parameters excluded by existing high energy scattering data.
Compared with earlier analyses, the present work constitutes a clear improvement, using all of the available data and allowing for variations in the PDFs associated with the introduction of the dark photon.

On the other hand, introducing the possibility of a dark photon yields a preference for the dark photon model of 6.5$\sigma$.
The inclusion of the dark photon also leads to a significant reduction in the size of the muon $g - 2$ anomaly.
We have tested the viability of this hypothesis against possible missing higher order uncertainties, as well as the treatment of power corrections and the lower cutoff applied on $Q^2$, with the result that none of these qualitatively change the conclusions.

In the future our analysis could be improved by systematically computing all observables at next-to-next-to-leading order (NNLO) instead of the next-to-leading order accuracy employed in the current analysis.
We plan to also implement different heavy-quark scheme for the discussion of charm and bottom quark production data.
The most important improvement would, however, be direct searches in the mass region suggested by our analysis.

This analysis suggests that studies of DIS data, in conjunction with other experimental constraints, provide a promising means of probing BSM physics. 
Given its significance, as well as the importance of precise nucleon PDFs as input in the determination of many physical quantities, including the $W$-boson mass~\cite{Gao:2022wxk} and weak couplings~\cite{PVDIS:2014cmd}, or searches for physics beyond the SM at the LHC, it is important to explore the effect which we have observed in more detail.

\section*{Acknowledgments}
We would like to acknowledge helpful discussions with Tom\'as Gonzalo, Gary Hill, Dipan Sengupta, Anthony Williams and Ross Young. 
This work was supported by the DOE contract No.~DE-AC05-06OR23177, under which Jefferson Science Associates, LLC operates Jefferson Lab; and by the University of Adelaide and the Australian Research Council through the ARC Centre of Excellence for Dark Matter Particle Physics (CE200100008) and Discovery Project DP180102209 (MJW). 
The work of NS was supported by the DOE, Office of Science, Office of Nuclear Physics in the Early Career Program.


\bibliographystyle{JHEP}
\bibliography{bibliography}

\end{document}